\documentclass[10pt,aps,prc,twocolumn,floatfix,showpacs,superscriptaddress]{revtex4-1} 
\usepackage{graphicx}
\usepackage{amsmath}
\usepackage{sidecap}
\usepackage{array}
\usepackage{color}
\usepackage{lineno}

\usepackage{ulem}

\def \pT {$p_{\rm{T}}$ }

\newcommand{\vtwo}{$v_{2}$}
\def \GeVc {\mbox{$\mathrm{GeV} / c$}}
\def \dEdx {\mbox{$\mathrm{d}E / \mathrm{d}x\ $}}

\begin{document}%

\title{Centrality dependence of identified particle elliptic flow in relativistic heavy ion collisions at $\sqrt{s_{NN}}=$ 7.7--62.4~GeV}

\date{\today}

	
\affiliation{AGH University of Science and Technology, Cracow 30-059, Poland}
\affiliation{Argonne National Laboratory, Argonne, Illinois 60439, USA}
\affiliation{Brookhaven National Laboratory, Upton, New York 11973, USA}
\affiliation{University of California, Berkeley, California 94720, USA}
\affiliation{University of California, Davis, California 95616, USA}
\affiliation{University of California, Los Angeles, California 90095, USA}
\affiliation{Central China Normal University (HZNU), Wuhan 430079, China}
\affiliation{University of Illinois at Chicago, Chicago, Illinois 60607, USA}
\affiliation{Creighton University, Omaha, Nebraska 68178, USA}
\affiliation{Czech Technical University in Prague, FNSPE, Prague, 115 19, Czech Republic}
\affiliation{Nuclear Physics Institute AS CR, 250 68 \v{R}e\v{z}/Prague, Czech Republic}
\affiliation{Frankfurt Institute for Advanced Studies FIAS, Frankfurt 60438, Germany}
\affiliation{Institute of Physics, Bhubaneswar 751005, India}
\affiliation{Indian Institute of Technology, Mumbai 400076, India}
\affiliation{Indiana University, Bloomington, Indiana 47408, USA}
\affiliation{Alikhanov Institute for Theoretical and Experimental Physics, Moscow 117218, Russia}
\affiliation{University of Jammu, Jammu 180001, India}
\affiliation{Joint Institute for Nuclear Research, Dubna, 141 980, Russia}
\affiliation{Kent State University, Kent, Ohio 44242, USA}
\affiliation{University of Kentucky, Lexington, Kentucky, 40506-0055, USA}
\affiliation{Korea Institute of Science and Technology Information, Daejeon 305-701, Korea}
\affiliation{Institute of Modern Physics, Lanzhou 730000, China}
\affiliation{Lawrence Berkeley National Laboratory, Berkeley, California 94720, USA}
\affiliation{Max-Planck-Institut fur Physik, Munich 80805, Germany}
\affiliation{Michigan State University, East Lansing, Michigan 48824, USA}
\affiliation{Moscow Engineering Physics Institute, Moscow 115409, Russia}
\affiliation{National Institute of Science Education and Research, Bhubaneswar 751005, India}
\affiliation{Ohio State University, Columbus, Ohio 43210, USA}
\affiliation{Institute of Nuclear Physics PAN, Cracow 31-342, Poland}
\affiliation{Panjab University, Chandigarh 160014, India}
\affiliation{Pennsylvania State University, University Park, Pennsylvania 16802, USA}
\affiliation{Institute of High Energy Physics, Protvino 142281, Russia}
\affiliation{Purdue University, West Lafayette, Indiana 47907, USA}
\affiliation{Pusan National University, Pusan 609735, Republic of Korea}
\affiliation{University of Rajasthan, Jaipur 302004, India}
\affiliation{Rice University, Houston, Texas 77251, USA}
\affiliation{University of Science and Technology of China, Hefei 230026, China}
\affiliation{Shandong University, Jinan, Shandong 250100, China}
\affiliation{Shanghai Institute of Applied Physics, Shanghai 201800, China}
\affiliation{State University Of New York, Stony Brook, NY 11794, USA}
\affiliation{Temple University, Philadelphia, Pennsylvania 19122, USA}
\affiliation{Texas A\&M University, College Station, Texas 77843, USA}
\affiliation{University of Texas, Austin, Texas 78712, USA}
\affiliation{University of Houston, Houston, Texas 77204, USA}
\affiliation{Tsinghua University, Beijing 100084, China}
\affiliation{United States Naval Academy, Annapolis, Maryland, 21402, USA}
\affiliation{Valparaiso University, Valparaiso, Indiana 46383, USA}
\affiliation{Variable Energy Cyclotron Centre, Kolkata 700064, India}
\affiliation{Warsaw University of Technology, Warsaw 00-661, Poland}
\affiliation{Wayne State University, Detroit, Michigan 48201, USA}
\affiliation{World Laboratory for Cosmology and Particle Physics (WLCAPP), Cairo 11571, Egypt}
\affiliation{Yale University, New Haven, Connecticut 06520, USA}
\affiliation{University of Zagreb, Zagreb, HR-10002, Croatia}

\author{L.~Adamczyk}\affiliation{AGH University of Science and Technology, Cracow 30-059, Poland}
\author{J.~K.~Adkins}\affiliation{University of Kentucky, Lexington, Kentucky, 40506-0055, USA}
\author{G.~Agakishiev}\affiliation{Joint Institute for Nuclear Research, Dubna, 141 980, Russia}
\author{M.~M.~Aggarwal}\affiliation{Panjab University, Chandigarh 160014, India}
\author{Z.~Ahammed}\affiliation{Variable Energy Cyclotron Centre, Kolkata 700064, India}
\author{I.~Alekseev}\affiliation{Alikhanov Institute for Theoretical and Experimental Physics, Moscow 117218, Russia}
\author{A.~Aparin}\affiliation{Joint Institute for Nuclear Research, Dubna, 141 980, Russia}
\author{D.~Arkhipkin}\affiliation{Brookhaven National Laboratory, Upton, New York 11973, USA}
\author{E.~C.~Aschenauer}\affiliation{Brookhaven National Laboratory, Upton, New York 11973, USA}
\author{G.~S.~Averichev}\affiliation{Joint Institute for Nuclear Research, Dubna, 141 980, Russia}
% ...

\collaboration{STAR Collaboration}\noaffiliation

\begin{abstract}
Elliptic flow ($v_{2}$) values for identified particles at midrapidity in Au + Au collisions measured by the STAR experiment in the Beam Energy Scan at the Relativistic Heavy Ion Collider at $\sqrt{s_{NN}}=$ 7.7--62.4~GeV are presented for three centrality classes. The centrality dependence and the data at $\sqrt{s_{NN}}=$ 14.5~GeV are new.  Except at the lowest beam energies we observe a similar relative $v_{2}$ baryon-meson splitting for all centrality classes which is in agreement within 15\% with the number-of-constituent quark scaling. The larger $v_{2}$ for most particles relative to antiparticles, already observed for minimum bias collisions, shows a clear centrality dependence, with the largest difference for the most central collisions. Also, the results are compared with A Multiphase Transport Model and fit with a Blast Wave model.
\end{abstract}

\pacs{25.75.Ld. 25.75.Nq} 

\maketitle

\section{Introduction}
\label{sec_intro}
The Beam Energy Scan (BES) program at the Relativistic Heavy Ion Collider (RHIC) facility was initiated in the year 2010 to study the Quantum Chromodynamics (QCD) phase diagram~\cite{Aggarwal:2010cw}. In the years 2010 and 2011 the STAR (Solenoidal Tracker at RHIC) experiment recorded Au + Au collisions at $\sqrt{s_{NN}}$ = 7.7, 11.5, 19.6, 27, 39, and 62.4~GeV. In the year 2014 data were recorded at 14.5~GeV. The results reported here are for a pseudorapidity range of $|\eta| < 1$. Recently published results from identified particle elliptic flow ($v_2$) in minimum bias (0\%--80\% centrality) collisions revealed an energy-dependent difference in elliptic flow between particles and antiparticles~\cite{Adamczyk:2013gv}. This difference is increasing with decreasing collision energy and is almost identical for all baryons.  It is larger for baryons than mesons. These observations attracted the attention of various theory groups, which tried to reproduce the results with different assumptions in their model calculations. (See Refs. [25-28] in Ref.~\cite{Adamczyk:2013gv}.) The most recent attempts are found in Ref.~\cite{Ivanov:2014zqa} which uses three-fluid dynamics, and Ref.~\cite{Hatta:2015era} which keeps the equilibration but varies the chemical potential. In this paper we present the energy and centrality dependence of identified particle elliptic flow. The new centrality dependence might be important for distinguishing between the different models or for improving their input parameters.

This paper is organized as follows. Section~\ref{sec_14.5} presents the recent minimum bias data at $\sqrt{s_{NN}}$ = 14.5~GeV. Section~\ref{sec_cent} presents the centrality and energy dependence of $v_{2}$ as a function of transverse kinetic energy $m_{\rm{T}}-m_{0}$. Section~\ref{sec_AMPT} shows a comparison with A Multiphase Transport Model (AMPT) calculation. In Sec.~\ref{sec_BW}, blast wave fits to the data are shown and the results for the transverse expansion velocity as a function of beam energy are discussed. A summary is presented in Sec.~\ref{sec_sum}. 

\section{14.5 GeV data}
\label{sec_14.5}

\begin{figure}[ht!]
\centering
\includegraphics[width=1.05\linewidth]{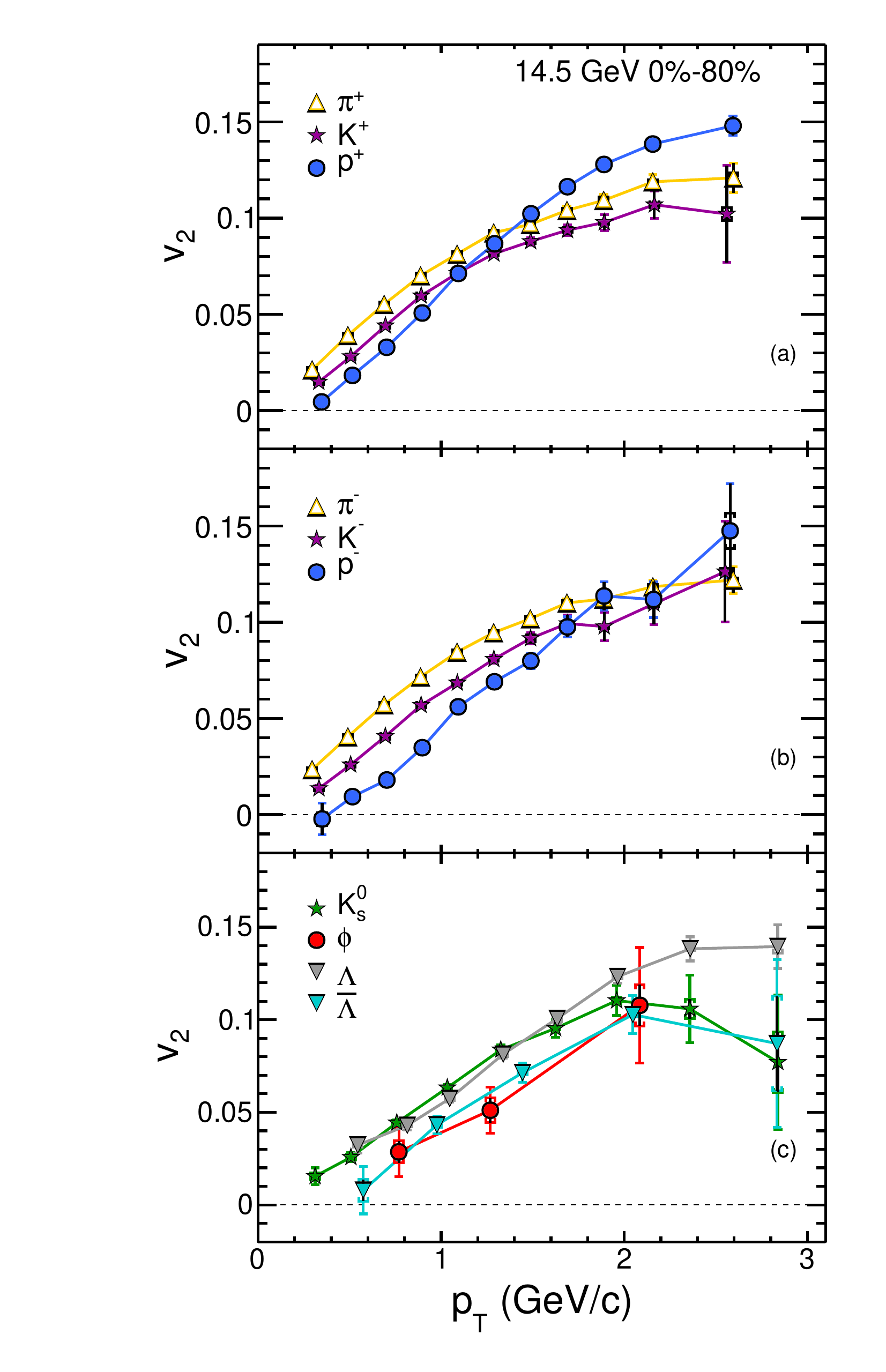} %
\caption{(Color online) Elliptic flow $v_2$ as a function of \pT for minimum bias data (0\%--80\% centrality) at $\sqrt{s_{NN}}$ = 14.5~GeV for identified particles. (a) positively charged particles. (b) negatively charged particles. (c) neutral particles. The systematic errors are shown by the short error bars with caps. The lines connect the points.}
\label{fig:14.5}
\end{figure}

\begin{figure}[ht!]
\centering
\includegraphics[width=1.05\linewidth]{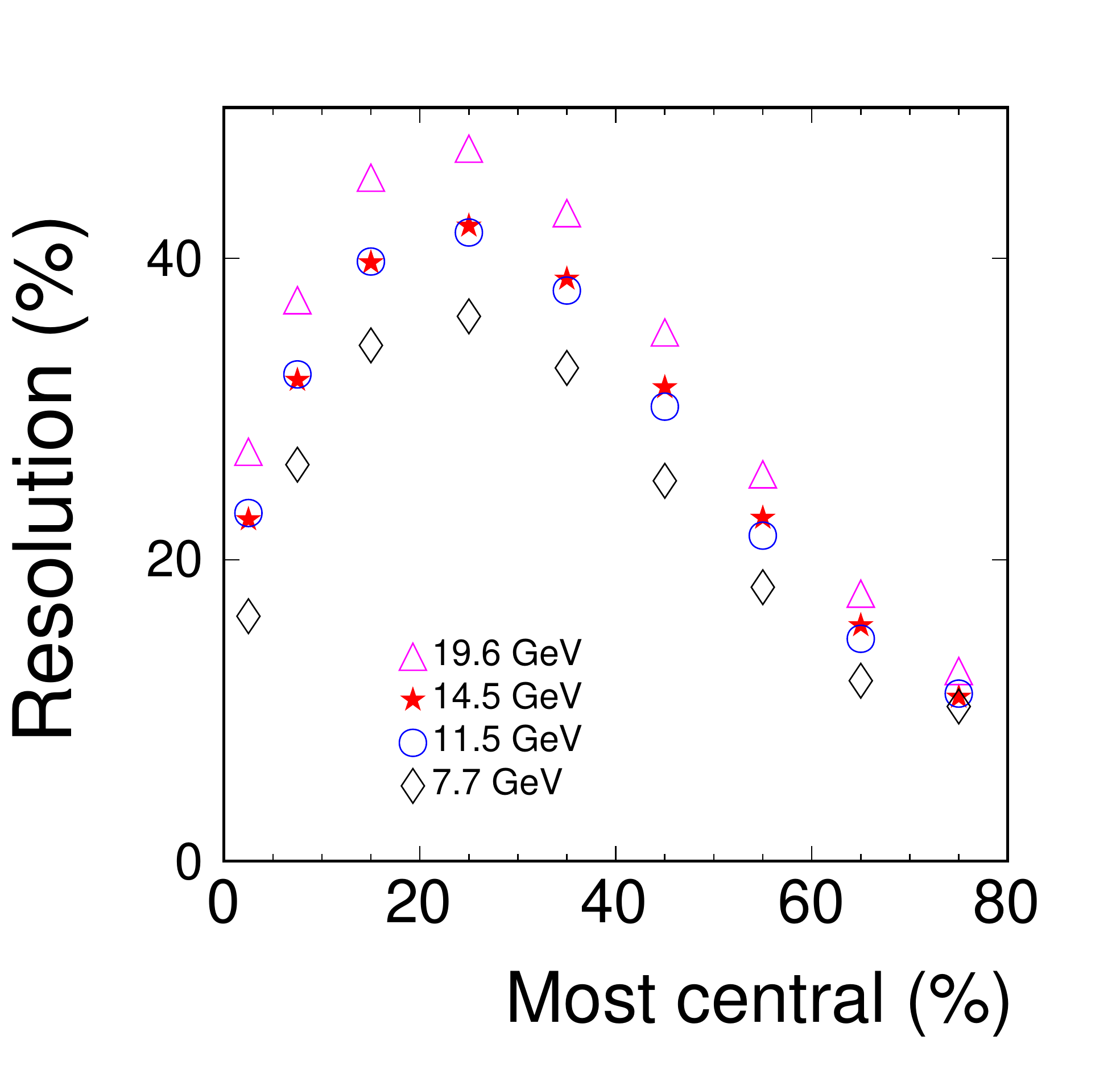} %
\caption{(Color online) The subevent plane resolution for several beam energies versus centrality with 5\% being the most central.}
\label{fig:res}
\end{figure}

The data obtained in 2014 at $\sqrt{s_{NN}}$ = 14.5~GeV were analyzed in the same way as the BES data at the other energies~\cite{Adamczyk:2012ku}. After a cut on the event vertex along the beam direction of $\pm$70 cm and a cut on the event vertex radial displacement from the mean of 1 cm, there were 17.5 M minimum-bias events available for data analysis. The centrality cuts on ``Reference Multiplicity", which is calculated with all reconstructed
particles within $|\eta| < 0.5$ and a distance-of-closest approach to the primary vertex smaller than 3 cm, were $>$200 particles for 0--10\% centrality, $>$59 and $<$200 particles for 10\%--40\% centrality, and $>$5 and $<$59 particles for 40\%--80\% centrality. The minimum bias results for all three centrality bins combined are shown in Fig.~\ref{fig:14.5}. The subevent plane resolution~\cite{Adamczyk:2013gw} is shown in Fig.~\ref{fig:res} compared to other beam energies from previous data sets in the BES. The 14.5~GeV resolution is close to the 11.5~GeV resolution because in 2014 there was additional material between the beam pipe and the Time Projection Chamber (TPC). This caused a lower multiplicity giving a slightly lower resolution than expected based on the other beam energies. 

\section{Centrality and Energy Dependence}
\label{sec_cent}
We present the transverse kinetic energy dependence of $v_{2}$ for 0\%--10\%, 10\%--40\% and 40\%--80\% central Au + Au collisions at $\sqrt{s_{NN}}$ = 7.7, 11.5, 14.5, 19.6, 27, 39, and 62.4~GeV. The analysis techniques used for particle identification, event plane reconstruction, and $v_{2}$ extraction are the same as the ones previously described~\cite{Adamczyk:2013gw}, and are summarized below. 

The identification of charged particles is based on a combination of momentum information, the specific energy loss \dEdx in the Time-Projection Chamber (TPC), and a required time-of-flight measurement with the Time-of-Flight detector (ToF). Charged pions and kaons can be easily distinguished up to 1.0~GeV/$c$ in transverse momentum, whereas at higher momenta the particle species start to significantly overlap. At higher \pT two-dimensional Gaussian fits in a combined $m^{2}$ vs. \dEdx plane were used to statistically extract the particle yield for $\pi^{\pm}$ and K$^{\pm}$ as a function of the relative angle to the reconstructed event plane angle $\Psi$.  For protons only one-dimensional Gaussian fits in $m^{2}$ were used to get the yields. The unstable particles K$_{s}^{0}$, $\phi$, $\Lambda$, $\Xi$, and $\Omega$, were reconstructed via the invariant mass technique. For weak decay particles, additional topological constraints~\cite{Adamczyk:2013gw} on the decay kinematics were applied to suppress background. The remaining combinatorial background was subtracted using the mixed event technique.
 
The event plane was reconstructed using charged particle tracks in the TPC. To suppress nonflow contributions we utilized the $\eta$-sub method, with an additional $\eta$-gap of $\pm$0.05 between the subevents and then averaged the results from the two subevents.  Recentering, $\phi$-weight, and shift techniques were applied for each $\eta$ hemisphere independently to flatten the event plane~\cite{Voloshin:2008dg}. The event plane resolution increases with $\sqrt{s_{NN}}$, with maxima as a function of centrality of 35\% at $\sqrt{s_{NN}}$ = 7.7~GeV and 50\% (not shown) at $\sqrt{s_{NN}}$ = 62.4~GeV. The systematic errors were estimated as in the previous publication~\cite{Adamczyk:2013gw}.

\begin{figure*}[ht!]
\centering
..
  \includegraphics[width=0.98\linewidth]{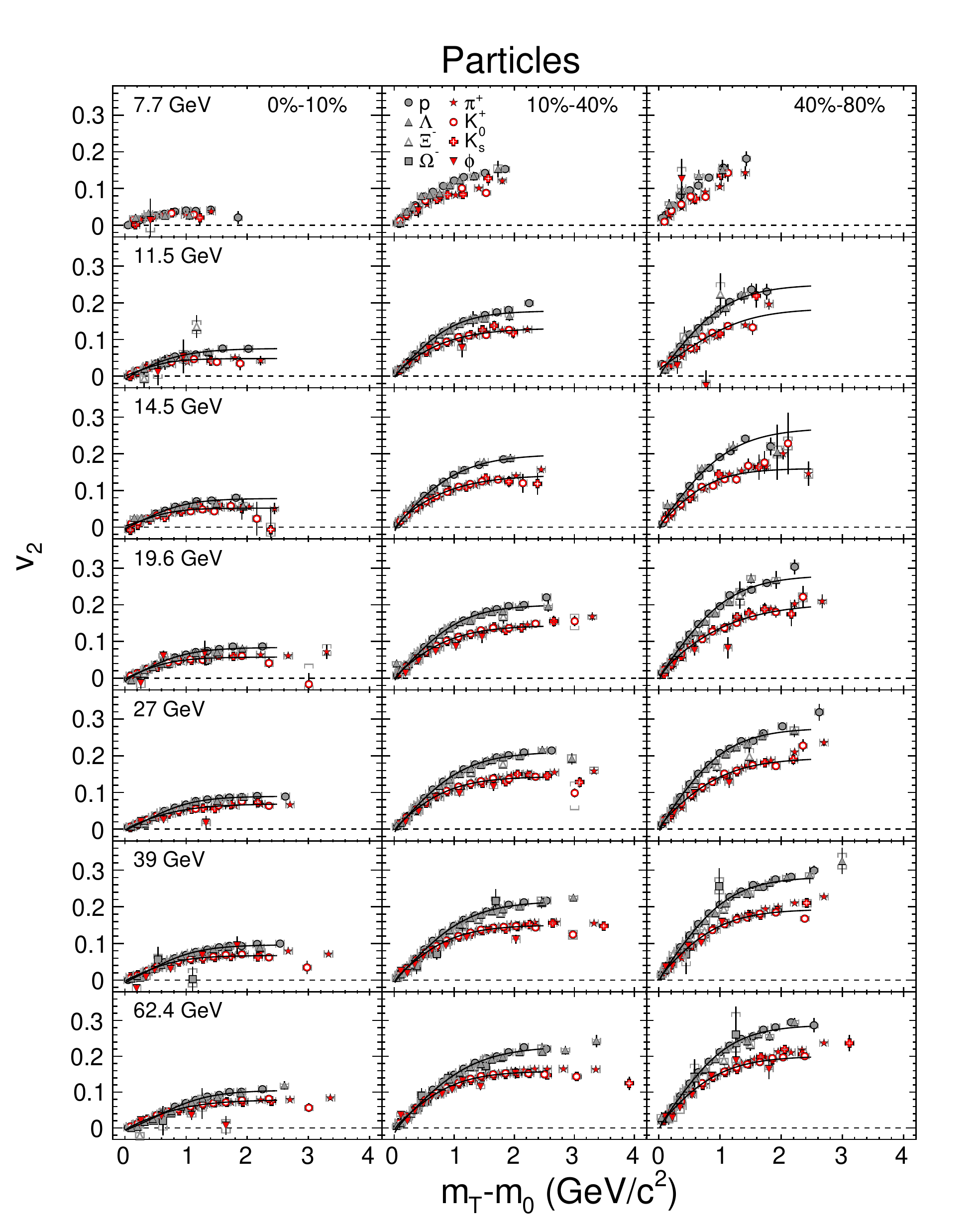}
  \caption{(Color online) The elliptic flow $v_{2}$ of identified particles ($\pi^{+}$, $K^{+}$, $K_{s}^{0}$, $p$, $\phi$, $\Lambda$, $\Xi^{-}$, $\Omega^{-}$) as a function of $m_{\rm{T}}-m_{0}$, for 0\%--10\%, 10\%--40\%, and 40\%--80\%  central Au + Au collisions at $\sqrt{s_{\rm{NN}}}$ = 7.7, 11.5, 14.5, 19.6, 27, 39, and 62.4~GeV. The lines show simultaneous fits to baryons and mesons with Eq. (\ref{Func_v2_fit}). The systematic errors are shown by the hooked error bars.}
\label{v2_mt_m0_pos}
\end{figure*}

Figure~\ref{v2_mt_m0_pos} shows $v_{2}$ vs. $m_{\rm{T}}-m_{0}$ of particles ($\pi^{+}$, $K^{+}$, $K^{0}_{s}$, $p$, $\phi$,$\Lambda$, $\Xi^{-}$, and $\Omega^{-}$) for  three centrality ranges of Au + Au collisions at $\sqrt{s_{NN}} = $ 7.7, 11.5, 14.5, 19.6, 27, 39, and 62.4~GeV. A splitting between baryons and mesons is observed at all energies and centralities except for 7.7~GeV central collisions. Here there are not enough events to allow a conclusion. All the $v_{2}$ values increase from central to peripheral collisions.

\begin{figure*}[ht!]
\centering
  \includegraphics[width=0.979\linewidth]{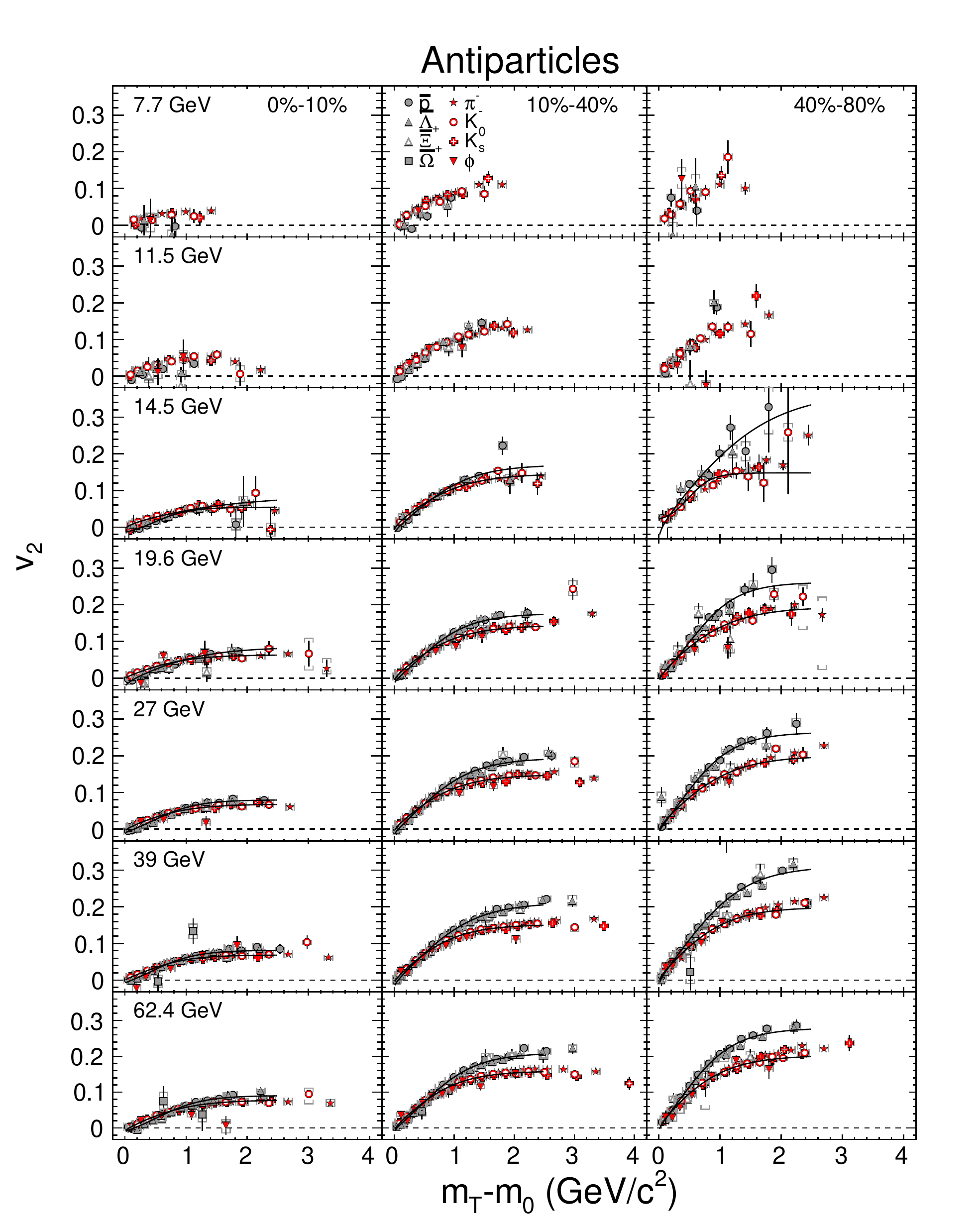}
  \caption{(Color online) The elliptic flow $v_{2}$ of identified antiparticles ($\pi^{-}$, $K^{-}$, $K_{s}^{0}$, $\bar{p}$, $\phi$, $\bar{\Lambda}$, $\bar{\Xi}^{+}$, $\bar{\Omega}^{+}$) as a function of $m_{\rm{T}}-m_{0}$, for 0\%--10\%, 10\%--40\%, and 40\%--80\%  central Au + Au collisions at $\sqrt{s_{\rm{NN}}}$ = 7.7, 11.5, 14.5, 19.6, 27, 39, and 62.4~GeV. The lines show simultaneous fits to baryons and mesons with Eq. (\ref{Func_v2_fit}). The systematic errors are shown by the hooked error bars.}
\label{v2_mt_m0_neg}
\end{figure*}

Figure~\ref{v2_mt_m0_neg} shows the energy and centrality dependence of $v_{2}$ vs. $m_{\rm{T}}-m_{0}$ but for antiparticles ($\pi^{-}$, $K^{-}$, $K^{0}_{s}$, $\bar{p}$, $\phi$, $\bar{\Lambda}$, $\bar{\Xi}^{+}$, and $\bar{\Omega}^{+}$). ($K^{0}_{s}$ and $\phi$ are plotted again since they are their own antiparticles.) The splitting between baryons and mesons is significant at 19.6~GeV and higher energies, and marginally significant at 14.5~GeV. There is no observed splitting for all centralities at 11.5~GeV and below. For these energies we are limited by the number of events and cannot draw a conclusion. For the $\phi$ meson at 14.5~GeV there were not enough events to plot the centrality dependence.

In both Figs.~\ref{v2_mt_m0_pos} and \ref{v2_mt_m0_neg}, for every particle species, energy, and centrality, $v_{2}$ increases with increasing $m_{\rm{T}}-m_{0}$. At $m_{\rm{T}}-m_{0}$ values larger than 1~GeV/$c^{2}$ an onset of $v_{2}$ saturation can be observed. For the most central 0\%--10\% collisions the absolute baryon-meson splitting is significantly smaller compared to more peripheral collisions, partly because the values are smaller making the absolute difference smaller.

\begin{figure}[ht!]
\centering
\includegraphics[width=0.98\linewidth]{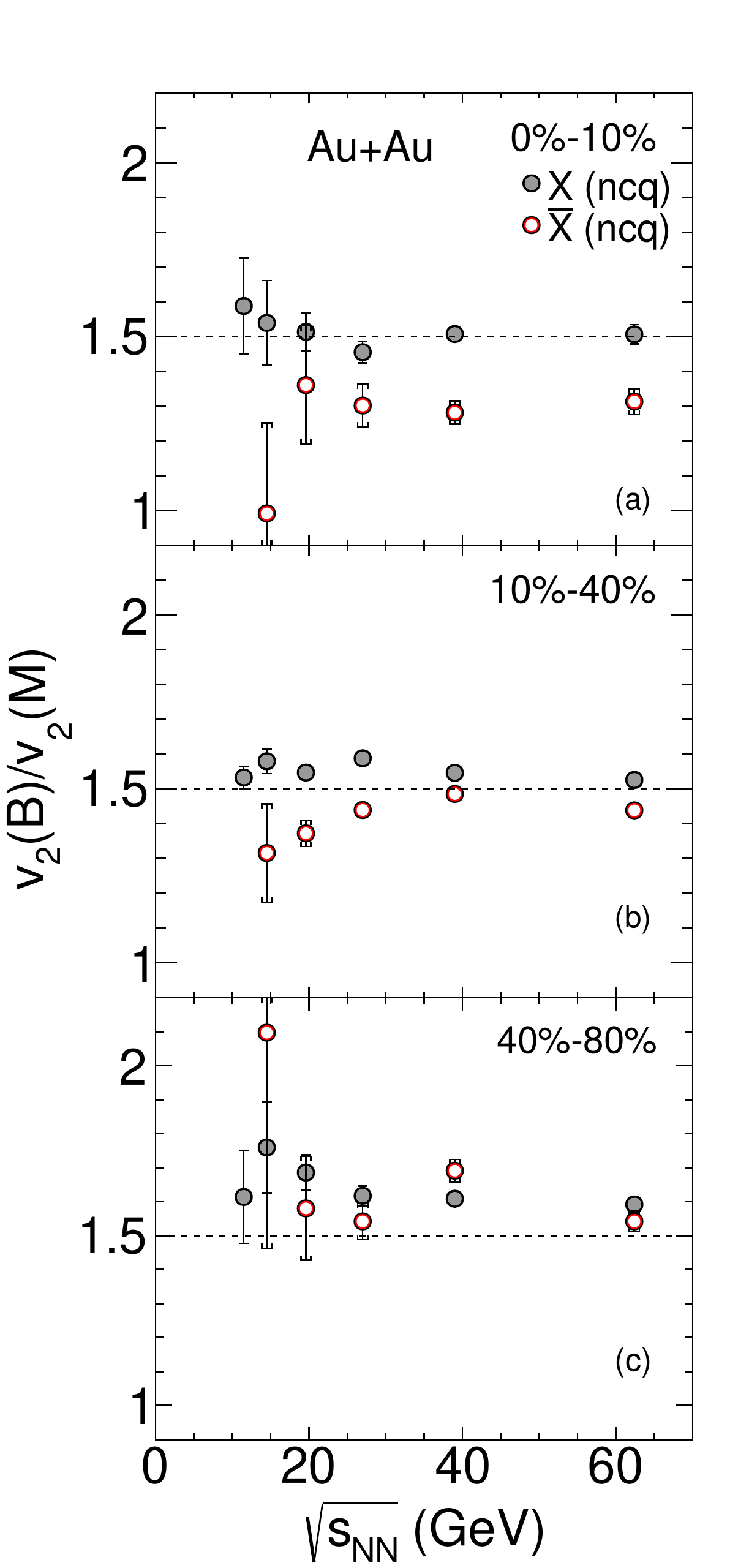} %
\caption{(Color online) The ratio of $v_{2}$ between baryons (B) and mesons (M) of particles ($\rm{X}$) and antiparticles (${\rm \overline{X}}$) as a function of $\sqrt{s_{NN}}$ for 0\%--10\%, 10\%--40\% and 40\%--80\% central Au + Au collisions. The values of baryons and mesons are taken from the fit lines in Figs.~\ref{v2_mt_m0_pos} and \ref{v2_mt_m0_neg} with Eq.~\ref{Func_v2_fit} at the appropriate values of $m_{\rm{T}}-m_{0}$. See text for details. The open points are for antiparticles and the closed points for particles.}
\label{v2B_v2M_sNN}
\end{figure}

\begin{figure}[ht!]
\centering
\includegraphics[width=0.98\linewidth]{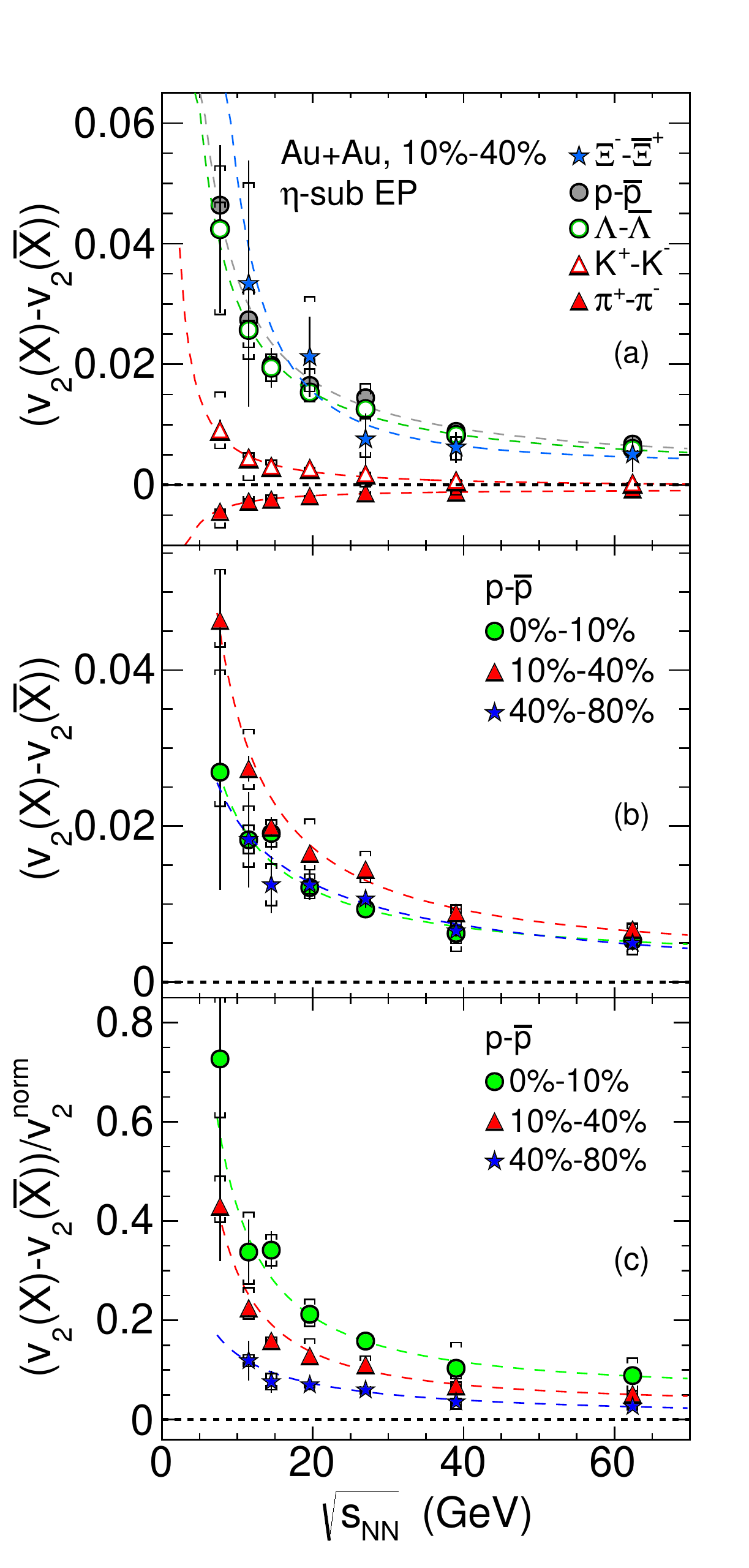} %
\caption{(Color online) (a) The difference in $v_{2}$ between particles ($\rm{X}$) and their corresponding antiparticles (${\rm \overline{X}}$) (see legend) as a function of $\sqrt{s_{\rm{NN}}}$ for 10\%--40\% central Au + Au collisions. (b) The difference in $v_{2}$ between protons and antiprotons as a function of $\sqrt{s_{\rm{NN}}}$ for 0\%--10\%, 10\%--40\% and 40\%--80\% central Au + Au collisions. (c) The relative difference. The systematic errors are shown by the hooked error bars. The dashed lines in the plot are fits with a power-law function.}
\label{fDiff_v2_sNN}
\end{figure}

To quantify the baryon and meson splitting and the scaling with the number of constituent quarks (NCQ), we fit the baryons (B) and mesons (M) separately using the function:
\begin{equation}
\label{Func_v2_fit}
	f_{v_{2}}(p_T,n) = \frac{an}{1+e^{-(p_{T}/n-b)/c}}-dn,
\end{equation}
where $a, b, c$, and $d$ are fit parameters and $n$ is the number of constituent quarks in the particle~\cite{Dong:2004ve}.
The ratio $v_{2}({\rm{B}})/v_{2}({\rm{M}}$) is calculated by the following steps.
First, we fit baryons with $n$=3 and mesons with $n$=2 using Eq.~\ref{Func_v2_fit} for particles and for antiparticles.
Second, we take the $v_{2}$ value from Eq.~\ref{Func_v2_fit} at $m_{\rm{T}}-m_{0} = 2 \ \rm{GeV/}$$c^2$ for baryons and at $m_{\rm{T}}-m_{0} = 2 \times (2/3) \ \rm{GeV/}$$c^2$ for mesons. That is because we want to compare the corresponding $v_{2}$ value after baryons and mesons are scaled by the number of constituent quarks. These \pT values were chosen to be above the hydro region but still where there were data for the lowest beam energy.
If there is a perfect NCQ scaling, the ratio $v_{2}\rm{(B)}/v_{2}\rm{(M)}$ should be equal to 1.5.
In Fig.~\ref{v2B_v2M_sNN}, we show this ratio as a function of beam energy for particles and antiparticles in three centrality bins. We can see from Fig.~\ref{v2B_v2M_sNN} the baryon-to-meson elliptic flow ratio for particles is higher than for antiparticles at all energies for 0\%--10\% and 10\%--40\% central collisions, but has no significant difference between particles and antiparticles for 40\%--80\%. The ratio for antiparticles shows a centrality dependence which is increasing from central to peripheral from about 1.3 to 1.6. But the ratio for particles does not show a significant centrality dependence. And there is no significant beam energy dependence for the ratio of both particles and antiparticles for the points plotted, except for antiparticles at 10\%--40\% centrality. In addition, we can see from the ratio that NCQ scaling holds for particles at centralities of 0\%--10\% and 10\%--40\%, but the ratio is slightly larger at 40\%--80\%.

In Fig.~\ref{fDiff_v2_sNN} upper panel, we show the difference in $v_2$ between particles ($\pi^{+}, K^{+}, p, \Lambda$, and $\Xi^{-}$) and their corresponding antiparticles ($\pi^{-}, K^{-}, \bar{p}, \bar{\Lambda}$, and $\bar{\Xi}^{+}$) for 10\%--40\% centrality. The difference is obtained by taking the average ratio in the measured \pT range as was done in Ref.~\cite{Adamczyk:2013gw}. The 10\%-40\% results are not very different from those obtained with minimum bias events shown previously~\cite{Adamczyk:2012ku}, but now are shown as a function of centrality in the middle panel for protons and antiprotons. In the lower panel the relative difference normalized by $v_{2}^{\rm norm}$, the proton elliptic flow at \pT = 1.5~GeV/$c$, shows a clear centrality dependence with a bigger effect for the more central collisions.

A systematic check has been carried out with the first-harmonic event plane
reconstructed by the two Beam-Beam Counters (BBC)~\cite{Whitten:2008zz,Bieser:2002ah} covering $3.3 < |\eta| < 5.0$.
The technical details are explained in Ref.~\cite{Agakishiev:2011id}. In the $\eta$-subevent method for $v_2\{\eta{\rm -sub}\}$ there is an $\eta$-gap of at least 0.3 between the observed event plane and the particles correlated to it in the opposite hemisphere. But using the BBCs this gap is at least 2.0 units of pseudorapidity. Possible systematic uncertainties arise from nonflow, $i.e.$ azimuthal correlations not related to the reaction plane orientation. These non-related correlations arise from resonances, jets, quantum statistics, and final-state interactions like Coulomb effects. They are suppressed by the use of a different harmonic for the event plane and the relatively large pseudorapidity gap between the STAR TPC and the BBC detectors~\cite{Poskanzer:1998yz,Agakishiev:2011id}. In practice, $v_2\{\rm BBC\}$ was measured in the following way:
\begin{equation}
	v_2\{{\rm BBC\}} = \langle \cos[2 \phi - \Psi_1 -\Psi_2]  \rangle / \langle \cos(\Psi_1 -\Psi_2) \rangle,
\end{equation}
where $\Psi_1$ and $\Psi_2$ are the first-harmonic subevent planes from the two BBC detectors.

\begin{figure}[ht!]
  \includegraphics[width=1.0\linewidth]{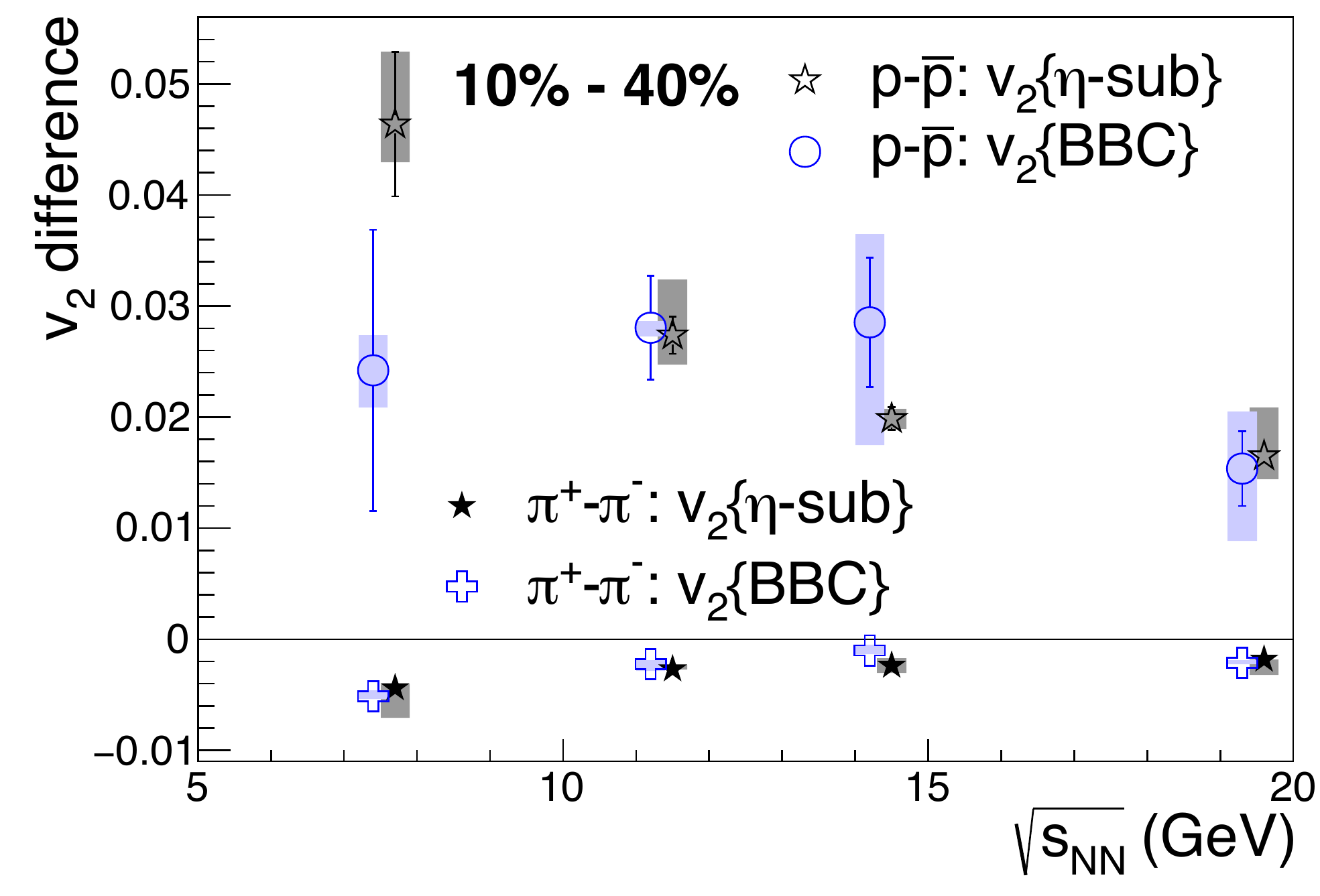}
  \caption{
    (Color online) The $v_2$ difference between protons and antiprotons (and between $\pi^+$ and $\pi^-$)
	for 10\%--40\% centrality Au+Au collisions at 7.7, 11.5, 14.5, and 19.6~GeV. The $v_2\{\rm BBC\}$ results were slightly
	shifted horizontally.
	}
\label{fig:BBC}
\end{figure}

The use of the first-harmonic event plane also reduces the event-by-event flow fluctuation
contribution compared with the $v_2\{\eta{\rm -sub}\}$ method in which the second-harmonic
event plane was used to calculate the second-harmonic anisotropy.
Figure~\ref{fig:BBC} presents a comparison between $v_2\{\rm BBC\}$ and $v_2\{\eta{\rm -sub}\}$,
in terms of the $v_2$ difference between protons and antiprotons (and between $\pi^+$ and $\pi^-$).
We focus on the center-of-mass energies below 20~GeV where the $v_2$ difference
between particles and antiparticles is most pronounced. For 10\%--40\% most central Au+Au collisions at 7.7, 11.5, 14.5, and 19.6~GeV, the results from the two methods are consistent with each other within the already quoted uncertainties.
This indicates that the $v_2$ difference is a robust observable and is not dominated
by nonflow or flow fluctuations.

\section{AMPT}
\label{sec_AMPT}

\begin{figure}[ht!]
\centering
\includegraphics[width=0.9\linewidth]{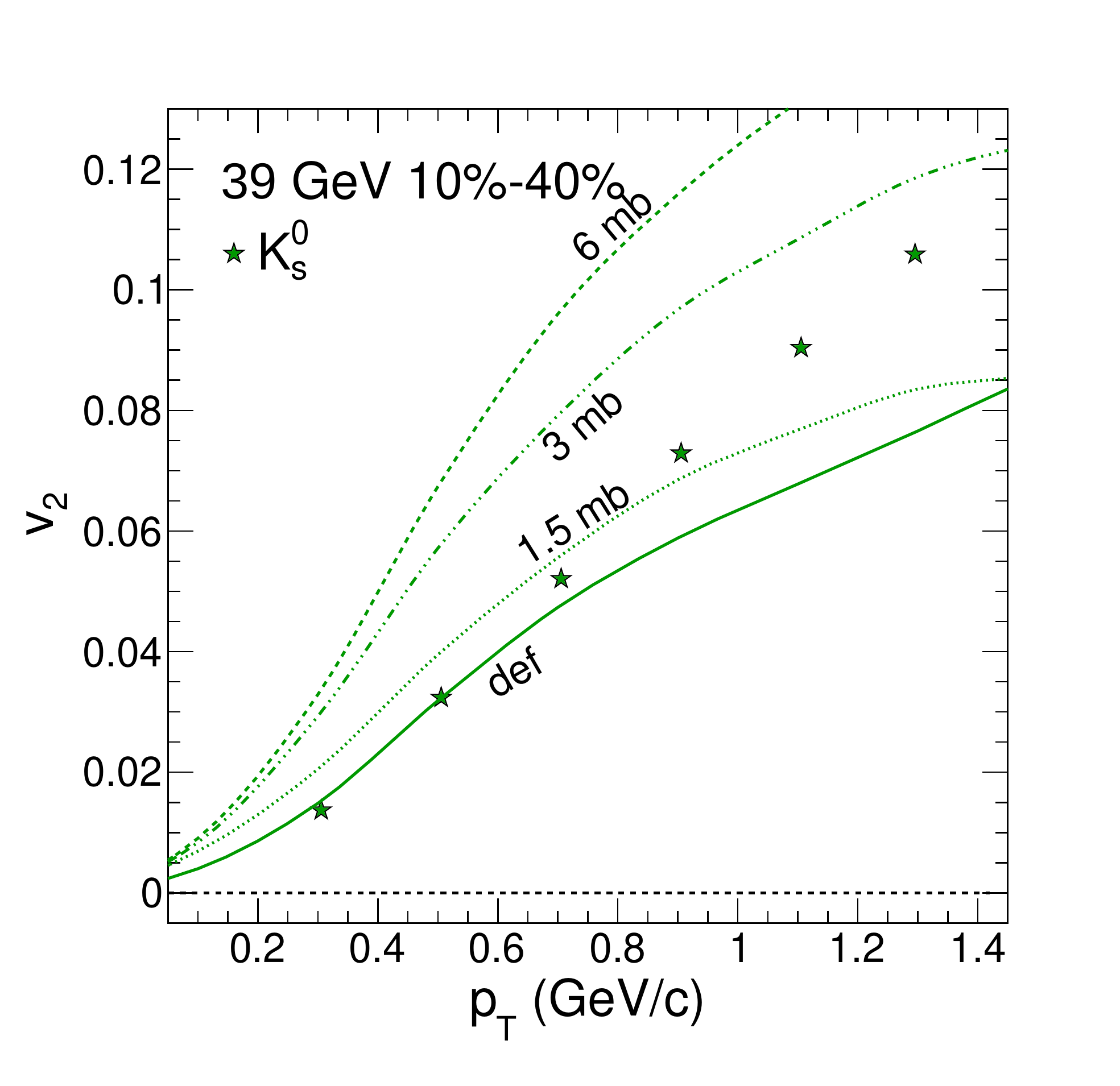} %
\caption{(Color online) Elliptic flow $v_2$ as a function of \pT for $K^0_s$ data at $\sqrt{s_{NN}}$ = 39~GeV for 10\%--40\% centrality. The curves are for AMPT default and AMPT string melting with cross sections of 1.5, 3.0, and 6.0~mb.}
\label{fig:AMPTK0}
\end{figure}

\begin{figure*}[ht!]
\centering
\includegraphics[width=0.7\linewidth]{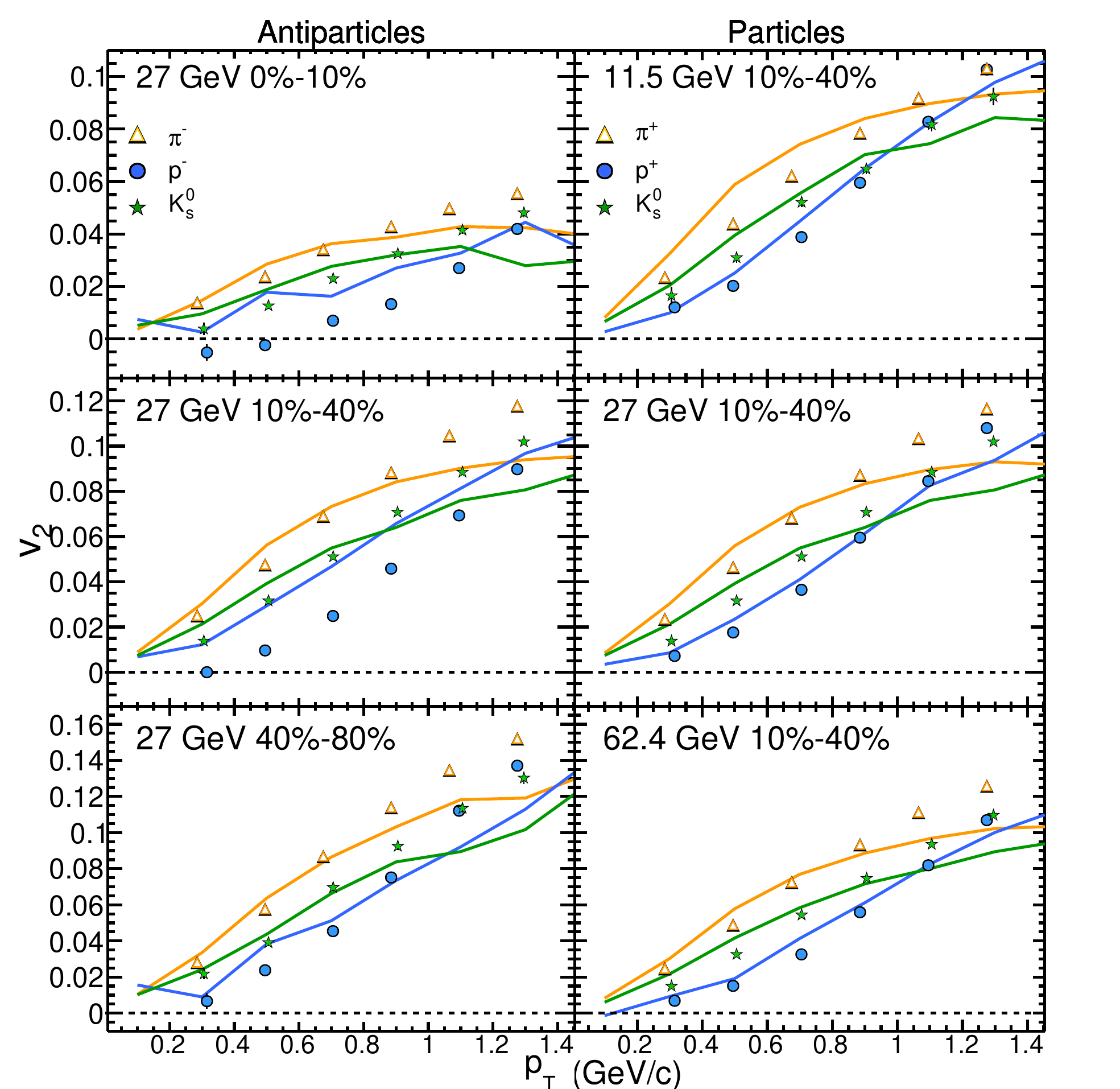} %
\caption{(Color online) Elliptic flow $v_2$ as a function of \pT for particles and antiparticles. The symbols show the experimental data. The error bars are mostly smaller than the points. The lines, with the same color code and the same order, show the AMPT string melting calculations with a cross section 1.5~mb. Antiparticles are on the left for three centrality bins. Particles are on the right for three beam energies.}
\label{fig:AMPT}
\end{figure*}

Calculations using AMPT were performed~\cite{Sun:2015pta}. The AMPT model is a transport model with four main components: the initial conditions, partonic interactions, conversion from the partonic to hadronic matter, and hadronic interactions~\cite{Lin:2004en}. It has two different versions to deal with different scenarios: default AMPT model and string melting AMPT model. The initial conditions are generated by HIJING (Heavy Ion Jet Interaction Generator) model~\cite{Gyulassy:1994ew,Wang:1990qp,Wang:1991hta}.
The HIJING model includes only two-body nucleon-nucleon interactions and generates minijets and excited strings through hard processes and soft processes separately. Excited strings are treated differently in the default and string melting  models. In the default  model, excited strings combine to form hadrons according to the Lund string fragmentation model, which then go through a hadronic interaction stage. In the string melting model, excited strings first convert to partons (melting) then have partonic interactions with the original soft partons.
The partonic interactions for both the default and string melting models are described by the ZPC (Zhang's Parton Cascade) model~\cite{Zhang:1997ej}. In the final stage of the ZPC model, partons in the default model recombine with parent strings and hadronize through the Lund string fragmentation model. However, in the string melting model, the hadronization of partons is described by a coalescence model. In both models after hadronization, the hadronic interactions are modelled by the ART (A Relativistic Transport) model~\cite{Li:1991mr,Li:1995pra}.

Approximately 10 to 20 million events were generated for each case for 0\%--80\% central Au+Au collisions at $\sqrt{s_{\rm NN}} = $ 11.5, 27, 39, and 62.4~GeV with the default  model (v1.25) and the string melting model (v2.25) with 3 different parton scattering cross sections (1.5~mb, 3~mb, and 6~mb). The same $\eta$-sub event method was used to calculate elliptic flow. Figure~\ref{fig:AMPTK0} shows $K^0_s$ data compared to $\sqrt{s_{NN}}$ = 39~GeV AMPT default and AMPT string melting with cross sections of 1.5, 3.0, and 6.0~mb. Although the shapes are not the same, the 1.5~mb curve seems to be the best compromise (see also Ref.~\cite{Sun:2015pta}). The curves with larger cross sections are all above the data points with deviations on the order of a factor 2 at \pT $<$ 2 \GeVc. Figure~\ref{fig:AMPT} shows comparisons of data with the AMPT string melting calculations with a cross section of 1.5~mb. The larger values of $v_2$ for protons compared to antiprotons can be seen in the middle panels for 27~GeV 10\%--40\%. Basic features of the data, like mass ordering and baryon-meson crossing at intermediate \pT, are well reproduced by AMPT. The calculations are furthermore in a reasonable quantitative agreement with the data for $K^0_s$ and protons, but deviate significantly for antiprotons in central and mid-central collisions. This shows that the particle-antiparticle difference, at least for protons, is not reproduced by AMPT. The pion \vtwo \  is similar at low \pT but systematically deviates to smaller values from the data at transverse momenta larger than 1 \GeVc.  

\begin{figure}[ht!]
\centering
\includegraphics[width=1.0\linewidth]{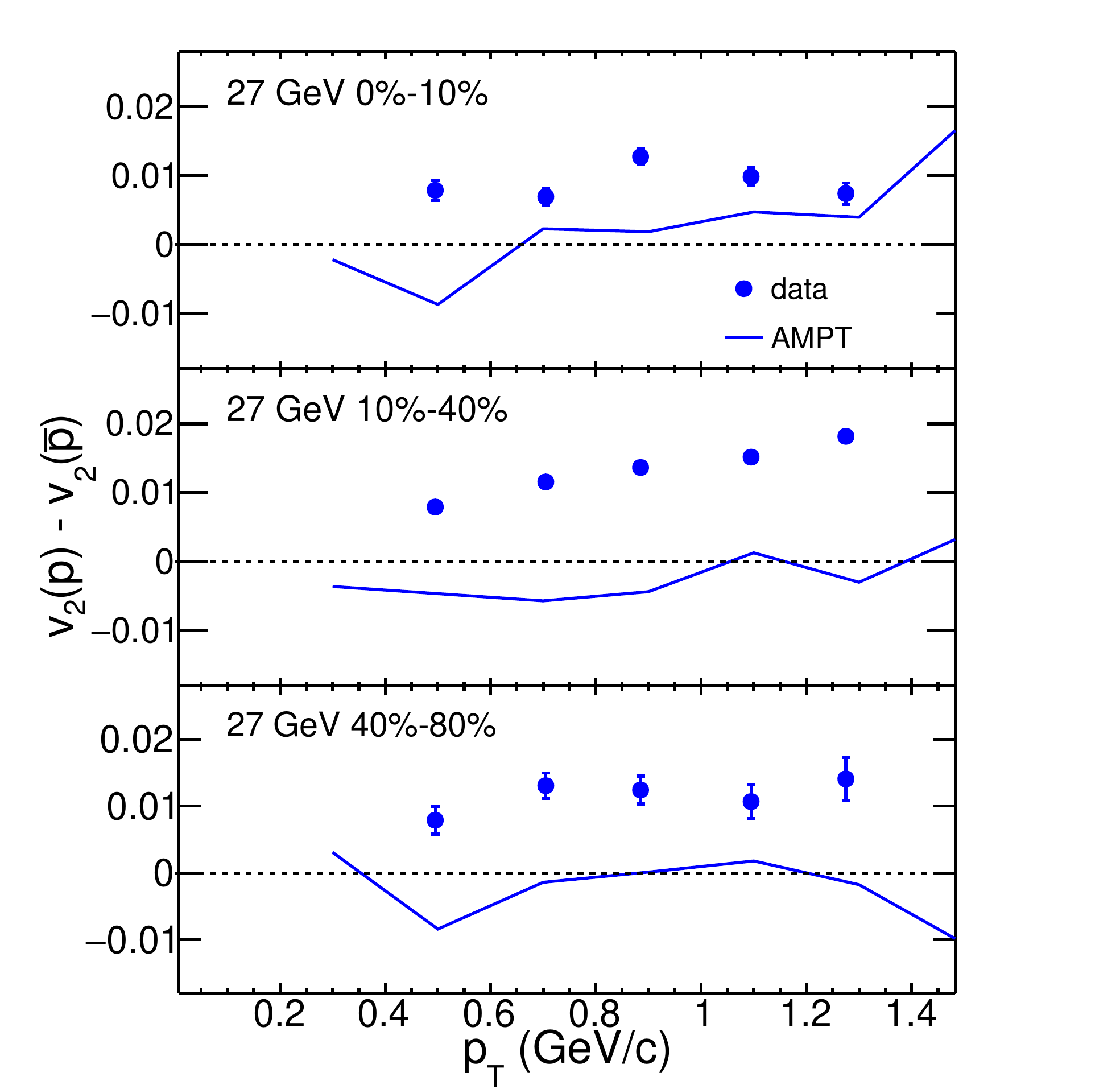} %
\caption{(Color online) Elliptic flow $v_2$ as a function of \pT for protons minus antiprotons at $\sqrt{s_{NN}}$ = 27~GeV for three centralities. The curves are for AMPT string melting with cross sections of 1.5~mb. The symbols are data.}
\label{fig:AMPTdif}
\end{figure}

Figure~\ref{fig:AMPTdif} shows the $v_2$ difference for protons minus antiprotons for $\sqrt{s_{NN}}$ = 27~GeV. It seems that there is little difference predicted by the AMPT calculations. AMPT does not explain the effect seen in the data. It was pointed out~\cite{Adamczyk:2013gw} that by including mean-field potentials~\cite{Xu:2012gf} in the hadronic phase of the AMPT model, the difference in elliptic flow between protons and antiprotons can be qualitatively reproduced, but then the charged kaon difference can not be reproduced.

\section{Blast Wave Fits}
\label{sec_BW}

\begin{figure*}[ht!]
\centering
\includegraphics[width=0.98\linewidth]{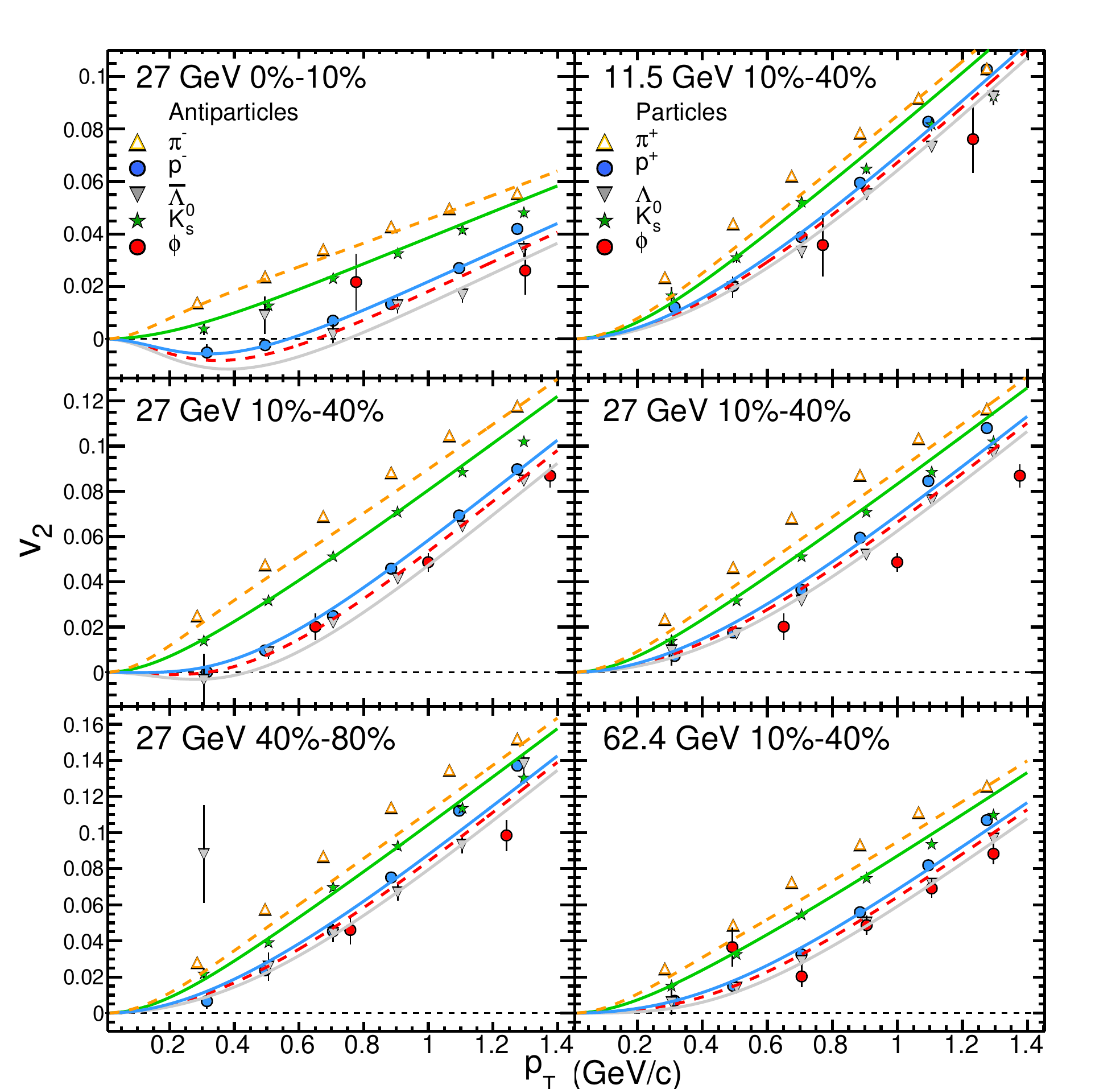}
\caption{(Color online) Elliptic flow $v_2$ as a function of \pT for particles and antiparticles. The data are shown by symbols. On the left side are Blast Wave fits (lines) for three centrality bins for $\sqrt{s_{NN}}=$ 27~GeV. On the right side are Blast Wave fits for three beam energies for 10\%--40\% centrality. Antiparticles are on the left, particles on the right. The lines are the same color and in the same order as the points. The dashed lines for $\pi$ and $\phi$ are not fits, but predictions based on the other fits. The error bars depict the combined statistical and systematic errors.}
\label{fig:BWfits}
\end{figure*}

In order to understand the hydrodynamic behaviour of $v_{2}(p_{\rm{T}})$ and its dependence on hadron mass and radial flow, we have applied a version of the ``blast wave" model~\cite{Adler:2001nb} which has four fit parameters: kinetic freeze-out temperature ($T$), transverse expansion rapidity ($\rho_{0}$), the momentum space variation in the azimuthal density ($\rho_{a}$), and the coordinate space variation in the azimuthal density ($s_{2}$). The blast wave equation we use is~\cite{Sun:2014rda}:

\begin{multline}
\label{eq:v2}
	v_{2}(p_{T}) = \\ \frac{\int_{0}^{2\pi}d\phi_{s} \cos(2\phi_{s})I_{2}[\alpha_{t}(\phi_{s})]K_{1}[\beta_{t}(\phi_{s})][1+2s_{2}\cos(2\phi_{s})]}{\int_{0}^{2\pi}d\phi_{s} I_{0}[\alpha_{t}(\phi_{s})] K_{1}[\beta_{t}(\phi_{s})][1+2s_{2}\cos(2\phi_{s})]}
\end{multline}
The $I_{0}$, $I_{2}$, and $K_{1}$ are modified Bessel functions, where $\alpha_{t}(\phi_{s}) = (p_{T}/T) \sinh[\rho(\phi_{s})]$, and $\beta_{t}(\phi_{s}) = (m_{T}/T)\cosh[\rho(\phi_{s})]$. The basic assumptions of this blast wave model are boost-invariant longitudinal expansion~\cite{Bjorken:1982qr} and freeze-out at constant temperature $T$ on a thin shell~\cite{Cooper:1974mv}, which expands with a transverse rapidity exhibiting a second harmonic azimuthal modulation given by $\rho(\phi_{s}) = \rho_{0} + \rho_{a}\cos2\phi_{s}$~\cite{Adler:2001nb}. In this equation, $\phi_{s}$ is the azimuthal angle in coordinate space; $\rho_{0}$ and $\rho_{a}$ are respectively the transverse expansion rapidity and the amplitude of its azimuthal variation. Secondly, $\beta = \tanh(\rho_{0})$, where $\beta$ is the transverse expansion velocity which is the velocity of the radial flow. Finally, $\beta_{a} = \tanh^{-1}(\rho_{a})$, where $\beta_a$ is the transverse expansion velocity second harmonic variation which is related to $v_2$. It needs to be noticed that the mass for different particle species enters in $m_{T}$ in $\beta_{t}(\phi_{s})$ only. When we do the simultaneous fits, which will be explained below, the only difference between the fits to different particle species is their mass.

We do blast wave fits for $v_{2}(p_{\rm{T}})$ for each centrality in the following way. 
First, we apply a cut on $m_{\rm{T}}-m_{0} < 0.9$~GeV to avoid the non-hydro region at high $p_{T}$. 
Second,	the fits for particles ($K^{+}, K^{0}_{s}, p$, and $\Lambda$) and antiparticles ($K^{-}, K^{0}_{s}, \bar{p}$, and $\bar{\Lambda}$) are separated, since we know that they have different behavior~\cite{Adamczyk:2013gv}. The $K^{0}_{s}$ and $\phi$ meson are plotted as both particles and antiparticles, since the antiparticles for $K^{0}_{s}$ and $\phi$ mesons are themselves. Third, pions are excluded from the fits since many pions come from feed-down from resonance decays~\cite{Brown:1991en}. This causes them not to have the proper shape for a Blast Wave equation fit. Also, $\phi$ mesons are not included in the fits because of their large error bars. Fourth, the fits are simultaneous fits which means that we use $v_{2}(p_{\rm{T}})$ of all of the species of particles or antiparticles to minimize the $\chi^{2}$ of the fit. We do not have spectra for most of the energies and therefore cannot use spectra to constrain the temperature. Instead we input a temperature in a reasonable range~\cite{NA49}. In this paper we choose $T = $ 120 MeV as the input, but will show also the results for 100 and 140 MeV.

In Fig.~\ref{fig:BWfits}, we show examples of the centrality and energy dependence of simultaneous blast wave fits for $K^{0}_{s}$, $p$, and $\Lambda$. The fits are done separately for particles and antiparticles. The dashed lines for $\pi$ and $\phi$ are not fits, but predictions based on the other fits. In the left side, we show the simultaneous blast wave fits for various centralities for antiparticles at 27~GeV.  We can see the splitting of different particle species is decreasing when we go from central to peripheral, which indicates the decreasing radial flow for antiparticles. In the right, we show the simultaneous blast wave fits for 10\%--40\% centrality at 11.5, 27, and 62.4~GeV for particles. We can see the splitting is slightly increasing with increasing energy, which indicates the increasing radial flow with increasing beam energy. If we compare the middle panel from the left and right sides, 10\%--40\% at 27~GeV for particles and antiparticles, we can see the splitting of antiparticles is larger than that of particles, which suggests the radial flow for antiparticles is larger than for particles. The pion predictions are somewhat low compared to data because the predictions do not include pions from resonance decay~\cite{Greco:2004ex}. It is worth noting that the $v_{2}$ values of $\phi$ meson are plotted at the same position for particles and antiparticles, but the predictions from blast wave (lower dashed lines in Fig.~\ref{fig:BWfits}) are different. The fits are different because they are dominated by protons and antiprotons, which are different. For most of the panels the agreement with the data is better with these fits. The $v_{2}$ values of $K^{0}_{s}$ are the same in both columns, and the $v_{2}$ of $K^{+}$ and $K^{-}$ (which are not shown here) are similar.

Although only examples of the fits are shown in Fig.~\ref{fig:BWfits}, all the fit parameters are shown in Table~\ref{table_fits}. At the lowest beam energy there were only enough data to fit the 10\%--40\% centrality. 
The goodness of fits were comparable to those reported in Ref.~\cite{Sun:2014rda}.
Without feed-down correction the $\chi^2$/ndf values
are only close to one at the lower energies, where the statistical
errors are of the order of the expected feed down effects.
At higher energies the error bars are much smaller. The
resulting $\chi^2$/ndf values rise up to a maximum of 35 for the
particle group at $\sqrt{s_{NN}}$ = 39 GeV, whereas they are below
1.5 for all energies when feed-down contributions{~\cite{Sun:2014rda} are
included into the error bars. For antiparticles the $\chi^2$/ndf values are
systematically lower compared to the particle group with a
maximum of 17, while they are about 1.5 with estimated
feed-down contributions taken into account.

\begin{figure}[ht!]
\centering
\includegraphics[width=0.98\linewidth]{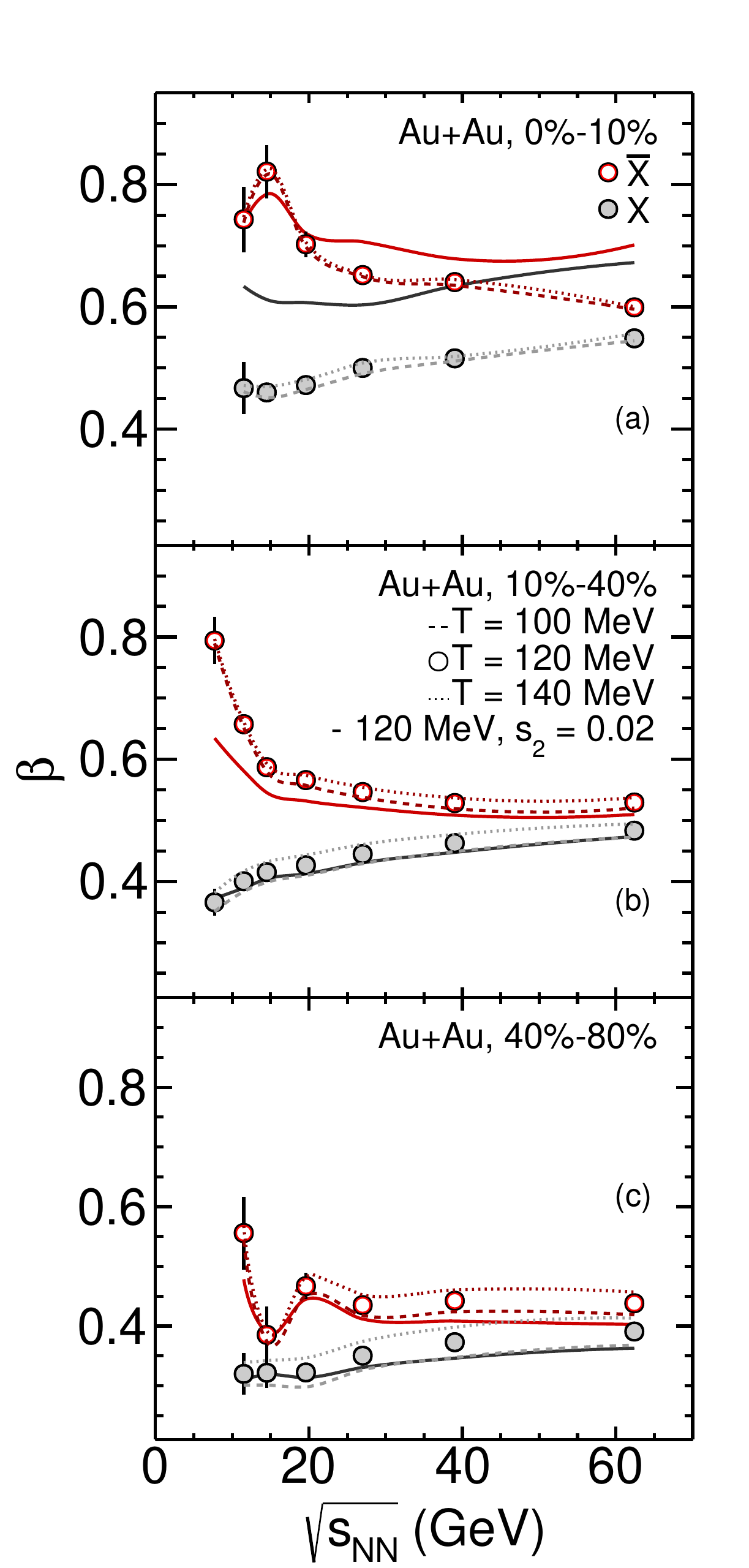} %
\caption{(Color online) The transverse radial velocity parameter $\beta$ as a function of $\sqrt{s_{NN}}$ for Au + Au collisions for three centralities and three assumed temperatures. The circles are for 120 MeV, open red circles for antiparticles and closed black circles for particles. The error bars seen at low beam energies depict the combined statistical and systematic errors. The solid red and black lines show the result when the $s_2$ parameter is fixed at 0.02 and the temperature held at 120 MeV.}
\label{fDiff_v2_sNN_muB}
\end{figure}

In Fig.~\ref{fDiff_v2_sNN_muB}, we show the transverse radial velocity parameter, which is extracted from the blast wave fits, as a function of beam energy for three centralities. We can see that at all three centralities the radial flow velocities for antiparticles are larger than for particles, and the difference in $\beta$ is generally increasing with decreasing energy.  This was already seen for minimum bias collisions~\cite{Sun:2014rda}, but now we also see it as a function of centrality. A large transverse radial velocity means that the $v_2(p_{\rm{T}})$ values are smaller because they are spread over a larger \pT range. The decrease in the difference between particles and antiparticles with increasing beam energy, suggests the radial flow velocities are becoming similar. Equal radial velocities have been observed at a beam energy of 200~GeV~\cite{Sun:2014rda}. We can see that the mean value of radial velocity for both particles and antiparticles is decreasing when we go from central to peripheral, which we have already seen from Fig.~\ref{fig:BWfits}. Another thing we have already seen from Fig.~\ref{fig:BWfits} is that the radial flow velocity is increasing with increasing beam energy for particles. To check if these trends are an artifact of the multi-parameter fitting procedure, we have fixed the $s_2$ parameter at 0.02 as shown in Fig.~\ref{fDiff_v2_sNN_muB}. It makes little difference for 10\%--40\% and 40\%--80\% centrality. However, for central collisions $\beta$ is larger with a smaller gap between particles and antiparticles. 

It is surprising to see a generally decreasing trend in $\beta$ for antiparticles with increasing beam energy.
We can speculate~\cite{Sun:2014rda} that at lower beam energy the antiparticles can only be produced at early time or not produced at all. Therefore, the produced antiparticles go through the whole expansion stage and get larger transverse expansion velocity than the particles which can be produced or transported in the latter stage. In addition, at lower collision energies, the absorption becomes important, especially for antibaryons. This effect also will lead to a higher value of mean \pT or, in the language of the Blast Wave fit, to a larger value of $\beta$. At higher beam energy, the antiparticles can be also produced in the latter stage of the evolution, and then only go through part of the expansion and get smaller transverse expansion velocity. At 14.5~GeV the bump for central collisions and the dip for peripheral collisions are probably statistical fluctuations plus some correlations with other parameters.

\begin{table*}[ht!]  
  \caption{\label{table_fits} Fit parameters $\rho_{0}$, $\rho_{a}$ and $s_{2}$ for the particle group ($\rm X$) and the antiparticle group ($\bar{\rm X}$) from Au + Au collisions at $\sqrt{s_{\rm NN}} = $ 7.7--62.4~GeV for three centralities.} 
  \footnotesize\rm  
  \begin{tabular}{ >{\centering\arraybackslash}m{0.7in}  >{\centering\arraybackslash}m{0.7in}  >{\centering\arraybackslash}m{0.70in} >{\centering\arraybackslash}m{0.70in} >{\centering\arraybackslash}m{0.70in} >{\centering\arraybackslash}m{0.70in} >{\centering\arraybackslash}m{0.70in} >{\centering\arraybackslash}m{0.70in} >{\centering\arraybackslash}m{0.70in}}  
    \hline \hline  
    $\rm centrality$&$\rm parameter$&$ 7.7\ \rm GeV $&$ 11.5\ \rm GeV $&$ 14.5\ \rm GeV $&$ 19.6\ \rm GeV $&$ 27\ \rm GeV $&$ 39\ \rm GeV $&$ 62.4\ \rm GeV $\\     \hline
    $0\%-10\%$&$\rho_{0} (\rm X)$&$ -- $&$ 0.51 \pm 0.05 $&$ 0.50 \pm 0.02 $&$ 0.51 \pm 0.01 $&$ 0.55 \pm 0.01 $&$ 0.57 \pm 0.01 $&$ 0.62 \pm 0.01 $\\ 
    $ $&$\rho_{0} (\bar{\rm X})$&$ -- $&$ 0.96 \pm 0.12 $&$ 1.16 \pm 0.13 $&$ 0.87 \pm 0.04 $&$ 0.78 \pm 0.02 $&$ 0.76 \pm 0.01 $&$ 0.69 \pm 0.02 $\\
    $ $&$\rho_{a} (\times 10^{-2} \ \rm X)$&$ -- $&$ 2.24 \pm 0.26 $&$ 2.36 \pm 0.09 $&$ 2.56 \pm 0.05 $&$ 2.46 \pm 0.08 $&$ 2.79 \pm 0.06 $&$ 2.97 \pm 0.04 $\\ 
    $ $&$\rho_{a} (\times 10^{-2} \ \bar{\rm X})$&$ -- $&$ 2.14 \pm 0.28 $&$ 1.54 \pm 0.26 $&$ 2.07 \pm 0.13 $&$ 2.46 \pm 0.09 $&$ 2.40 \pm 0.06 $&$ 2.98 \pm 0.08 $\\
    $ $&$s_{2} (\times 10^{-2} \ \rm X)$&$ -- $&$ 0.03 \pm 3.37 $&$ 0.00 \pm 1.10 $&$ 0.00 \pm 0.60 $&$ 0.52 \pm 0.18 $&$ 0.12 \pm 0.14 $&$ 0.00 \pm 0.64 $\\
    $ $&$s_{2} (\times 10^{-2} \ \bar{\rm X})$&$ -- $&$ 2.09 \pm 0.57 $&$ 2.32 \pm 0.35 $&$ 1.81 \pm 0.22 $&$ 1.31 \pm 0.17 $&$ 1.53 \pm 0.10 $&$ 0.37 \pm 0.19 $\\
    \hline  

    $10\%-40\%$&$\rho_{0} (\rm X)$&$ 0.38 \pm 0.03 $&$ 0.42 \pm 0.01 $&$ 0.44 \pm 0.01 $&$ 0.46 \pm 0.01 $&$ 0.48 \pm 0.00 $&$ 0.50 \pm 0.00 $&$ 0.53 \pm 0.00 $\\ 
    $ $&$\rho_{0} (\bar{\rm X})$&$ 1.08 \pm 0.10 $&$ 0.79 \pm 0.03 $&$ 0.67 \pm 0.02 $&$ 0.64 \pm 0.01 $&$ 0.61 \pm 0.01 $&$ 0.59 \pm 0.00 $&$ 0.59 \pm 0.01 $\\
    $ $&$\rho_{a} (\times 10^{-2} \ \rm X)$&$ 4.39 \pm 0.28 $&$ 4.41 \pm 0.14 $&$ 4.55 \pm 0.15 $&$ 4.47 \pm 0.08 $&$ 4.55 \pm 0.05 $&$ 4.75 \pm 0.04 $&$ 5.19 \pm 0.05 $\\
    $ $&$\rho_{a} (\times 10^{-2} \ \bar{\rm X})$&$ 3.47 \pm 0.36 $&$ 4.22 \pm 0.18 $&$ 4.48 \pm 0.17 $&$ 4.62 \pm 0.08 $&$ 4.80 \pm 0.05 $&$ 5.05 \pm 0.04 $&$ 5.24 \pm 0.05 $\\ 
    $ $&$s_{2} (\times 10^{-2} \ \rm X)$&$ 1.56 \pm 0.72 $&$ 2.66 \pm 0.31 $&$ 2.64 \pm 0.33 $&$ 2.94 \pm 0.16 $&$ 2.93 \pm 0.11 $&$ 2.82 \pm 0.08 $&$ 2.45 \pm 0.10 $\\
    $ $&$s_{2} (\times 10^{-2} \ \bar{\rm X})$&$ 5.02 \pm 0.70 $&$ 3.87 \pm 0.33 $&$ 3.19 \pm 0.32 $&$ 3.03 \pm 0.16 $&$ 2.81 \pm 0.11 $&$ 2.67 \pm 0.08 $&$ 2.71 \pm 0.10 $\\
    \hline 

    $40\%-80\%$&$\rho_{0} (\rm X)$&$ -- $&$ 0.33 \pm 0.04 $&$ 0.33 \pm 0.03 $&$ 0.33 \pm 0.02 $&$ 0.37 \pm 0.01 $&$ 0.39 \pm 0.01 $&$ 0.41 \pm 0.01 $\\
    $ $&$\rho_{0} (\bar{\rm X})$&$ -- $&$ 0.63 \pm 0.09 $&$ 0.41 \pm 0.06 $&$ 0.51 \pm 0.03 $&$ 0.47 \pm 0.02 $&$ 0.48 \pm 0.01 $&$ 0.47 \pm 0.01 $\\
    $ $&$\rho_{a} (\times 10^{-2} \ \rm X)$&$ -- $&$ 5.15 \pm 0.75 $&$ 5.93 \pm 0.67 $&$ 4.85 \pm 0.41 $&$ 5.13 \pm 0.22 $&$ 4.96 \pm 0.14 $&$ 5.58 \pm 0.17 $\\
    $ $&$\rho_{a} (\times 10^{-2} \ \bar{\rm X})$&$ -- $&$ 4.69 \pm 0.79 $&$ 7.09 \pm 0.28 $&$ 5.90 \pm 0.39 $&$ 5.75 \pm 0.23 $&$ 5.77 \pm 0.14 $&$ 5.92 \pm 0.17 $\\ 
    $ $&$s_{2} (\times 10^{-2} \ \rm X)$&$ -- $&$ 5.60 \pm 2.20 $&$ 3.87 \pm 1.87 $&$ 7.66 \pm 1.30 $&$ 6.91 \pm 0.59 $&$ 7.35 \pm 0.35 $&$ 6.39 \pm 0.39 $\\ 
    $ $&$s_{2} (\times 10^{-2} \ \bar{\rm X})$&$ -- $&$ 4.93 \pm 1.22 $&$ 0.00 \pm 1.52 $&$ 3.29 \pm 0.79 $&$ 3.83 \pm 0.48 $&$ 4.79 \pm 0.27 $&$ 5.31 \pm 0.34 $\\ 
    \hline \hline 
  \end{tabular}  
\end{table*} 

\section{Summary}
\label{sec_sum}

For 14 identified particles ($\pi^{-}$, $\pi^{+}$, $K^{-}$, $K^{+}$, $K^{0}_{s}$, $p$, $\bar{p}$, $\phi$, $\Lambda$, $\bar{\Lambda}$, $\Xi^{-}$, $\bar{\Xi}^{+}$, $\Omega^{-}$, and $\bar{\Omega}^{+}$), we have measured the elliptic flow $v_{2}$ for Au + Au collisions for seven beam energies ($\sqrt{s_{NN}}$ = 7.7, 11.5, 14.5, 19.6, 27, 39, and 62.4~GeV), and three centralities (0\%--10\%, 10\%--40\%, and 40\%--80\%). The Baryon-Meson splitting at intermediate \pT is in reasonable agreement with NCQ scaling for all energies and centralities reported here. The $v_2$ of baryons is larger than for antibaryons for all beam energies, and the relative increase for protons compared to antiprotons (see Fig.~\ref{fDiff_v2_sNN} (c)) is larger for central collisions.

AMPT calculations with string melting with a 1.5~mb partonic cross section do not explain the proton-antiproton difference.

With a Blast Wave model we have fit the results for particles ($K^{+}$, $K^{0}_{s}$, $p$, $\Lambda$) and antiparticles ($K^{-}$, $K^{0}_{s}$, $\bar{p}$, $\bar{\Lambda}$) separately with three blast wave parameters ($\rho_0, \rho_a$, and $s_2$). The significant parameter which changes the most with beam energy is the transverse radial velocity ($\beta$) which comes from $\rho_0$. Its value is much larger for antiparticles than particles, but the difference decreases with increasing beam energy. It is also larger for central collisions than peripheral collisions. The behavior of this transverse radial flow parameter quantifies the $v_{2}$ particle-antiparticle difference observed above and published previously for minimum bias collisions~\cite{Adamczyk:2013gv}. 

\section{Acknowledgements}

We thank the RHIC Operations Group and RCF at BNL, the NERSC Center at LBNL, the KISTI Center in
Korea, and the Open Science Grid consortium for providing resources and support. This work was 
supported in part by the Office of Nuclear Physics within the U.S. DOE Office of Science,
the U.S. NSF, the Ministry of Education and Science of the Russian Federation, NNSFC, CAS,
MoST and MoE of China, the National Research Foundation of Korea, 
GA and MSMT of the Czech Republic, FIAS of Germany, DAE, DST, and UGC of India, the National
Science Centre of Poland, National Research Foundation, the Ministry of Science, Education and 
Sports of the Republic of Croatia, and RosAtom of Russia.


\end{document}